\documentclass[epsfig,12pt]{article}
\usepackage{epsfig,amssymb,euscript,bbold}
\usepackage{amsmath}
\allowdisplaybreaks
\newcommand{\be}{\begin{equation}}
\newcommand{\ee}{\end{equation}}
\newcommand{\bea}{\begin{eqnarray}}
\newcommand{\eea}{\end{eqnarray}}
\newcommand{\ba}{\begin{array}}
\newcommand{\p}[1]{(\ref{#1})}
\newcommand{\ea}{\end{array}}

\def\bbox{{\,\lower0.9pt\vbox{\hrule \hbox{\vrule height 0.2 cm
\hskip 0.2 cm \vrule height 0.2 cm}\hrule}\,}}
\newcommand{\dsl}{\pa \kern-0.5em /}

\newcommand{\nn}{\nonumber \\}

\def\ds{\raise.15ex\hbox{/}\kern-.57em\partial}
\def\Ds{\,\raise.15ex\hbox{/}\mkern-13.5mu D}

\newcommand{\e}{\epsilon}

\newcommand{\bR}{\mathbb{R}}

\font\mybb=msbm10 at 10pt
\def\bb#1{\hbox{\mybb#1}}

\def\bR {\bb{R}}

\begin{document}



\begin{titlepage}

\vfill

\begin{flushright}
hep-th/0406188\\
\end{flushright}

\vfill

\begin{center}

\baselineskip=16pt

{\Large\bf A deformation of $AdS_5\times S^5$}

\vskip 1.3cm

Jerome P. Gauntlett$^{1*}$, Jan B. Gutowski$^{2}$ and Nemani V. Suryanarayana$^{1}$

\vskip 1cm

{\small{\it $^1$Perimeter Institute for Theoretical
Physics\\ Waterloo, ON, N2J 2W9, Canada\\ E-mail:
jgauntlett, vnemani@perimeterinstitute.ca\\}}
\vskip .6cm
{\small{\it
$^{2}$Mathematical Institute, Oxford University, \\Oxford, OX1 3LB, U.K.\\
E-mail: gutowski@maths.ox.ac.uk \\}}

\end{center}

\vfill

\begin{center}
\textbf{Abstract}
\end{center}
We analyse a one parameter family of supersymmetric
solutions of type IIB supergravity that includes $AdS_5\times S^5$.
For small values of the parameter the solutions are causally well-behaved, but beyond
a critical value closed timelike curves (CTC's) appear. The solutions are holographically
dual to ${\cal N}=4$
supersymmetric Yang-Mills theory on a non-conformally flat background with
non-vanishing $R$-currents. We compute the holographic energy-momentum tensor for the
spacetime and show that it remains finite even when the CTC's appear.
The solutions, as well as the uplift of some recently discovered $AdS_5$
black hole solutions, are shown to preserve precisely two supersymmetries.
\begin{quote}

\end{quote}

\vfill

\vfill \vskip 5mm \hrule width 5.cm \vskip 5mm
{\small{\noindent $^*$On leave from: Blackett Laboratory, Imperial
  College, London, SW7 2BZ, U.K.\\
  }}

\end{titlepage}

\makeatletter
\renewcommand{\theequation}{\thesection.\arabic{equation}}
\@addtoreset{equation}{section} \makeatother


\section{Introduction}

A powerful way to generalise the basic examples of
the $AdS/CFT$ correspondence in string and M-theory \cite{mal} is to use
gauged supergravities. For example, the bosonic sector
of minimal gauged supergravity in D=5 \cite{gun}
can be obtained via a consistent truncation of the Kaluza-Klein reduction of type
IIB supergravity on a five-sphere \cite{cejm},
which implies that any bosonic solution of
the gauged supergravity can be uplifted to obtain a solution of type IIB supergravity.
The $AdS_5$ vacuum solution uplifts to the $AdS_5\times S^5$ solution, which is
dual to four-dimensional
${\cal N}=4$ super-symmetric Yang-Mills theory, while non-vacuum solutions
uplift to solutions containing a deformed five-sphere, which
can correspond to various deformations of this CFT.

In \cite{gaungut} a one-parameter family of supersymmetric
solutions of D=5 minimal gauged
supergravity were constructed, which includes the $AdS_5$ vacuum solution
when the parameter is set to zero.
We will show here that the family of solutions are
asymptotically locally $AdS_5$, and hence the
corresponding uplifted type IIB solutions
can be interpreted as being dual to
${\cal N}=4$ supersymmetric Yang-Mills theory living on
the non-trivial conformal boundary, which we show is not conformally flat.
Since the gauged supergravity solutions have non-vanishing $U(1)$ gauge-fields,
this implies that $R$-symmetry currents are also non-vanishing in the
dual picture. Note that this interpretation was also discussed in \cite{BK},
where the solutions were further generalised (see also \cite{Behrndt:2003gc}).

An interesting feature of the solutions of \cite{gaungut} is
that closed time-like curves (CTCs) appear
when the parameter is larger than a critical value.
There have been several recent investigations aiming
to understand the role, if any, of CTC's in string theory. This activity was
catalysed by the discovery of the supersymmetric G\"odel solution of
minimal {\it ungauged} supergravity in D=5 \cite{gghpr}.
These solutions are homogeneous and preserve twenty supersymmetries
when uplifted to provide a solution of D=11 supergravity.
In \cite{Boyda:2002ba} it was suggested that the physics of these
spacetimes might be encoded holographically in observer dependent screens of
the type discussed in \cite{bousso}.
Somewhat related to this idea is the possibility that only a region of the
G\"odel spacetime makes sense in string theory: the rest of the spacetime,
including the CTCs, should be excised and replaced by another spacetime in order to get a
physically sensible background \cite{Dyson:2003zn,Drukker:2003sc,Drukker:2004zm,Gimon:2004if}
(for related work see, for example, \cite{Jarv:2002wu}-\cite{Fiol:2003yq}).
However, it is fair to say
that the matter is not yet settled.

It is natural to wonder if standard $AdS/CFT$  holography can provide a
new perspective on CTCs. For example, it would be interesting if there
are
supersymmetric solutions of string/M-theory that are asymptotically locally
$AdS$ with CTCs confined to the bulk. Such solutions might then
be dual to a CFT on the boundary which could then provide a precise description, in principle,
of string/M-theory propagating on the spacetime with CTCs.
In any case, we will show here that the solutions of \cite{gaungut} are not in this class: for values of the
parameter when the CTC's appear in the bulk, they also appear on the boundary.

Nevertheless, one might boldly assume that the duality is still valid
for our solutions for supercritical values of the parameter.
The supergravity solution should then be dual to ${\cal N}=4$ SYM on
a background containing CTCs. Now from intuition garnered from analysing quantum fields
on curved spacetimes, one might expect that the energy-momentum tensor of the field
theory to diverge on such a spacetime. Indeed, the energy-momentum tensor of a scalar
field propagating on a one-parameter family of solutions that includes
the four-dimensional G\"odel spacetime \cite{Reboucas:hn}
was calculated using zeta-function regularisation
in \cite{Huang:zi}. As the parameter is varied, it was shown that the
energy-momentum tensor becomes divergent precisely at the onset of CTCs.

Here we will calculate the holographic energy-momentum (EM) tensor for
our family of solutions using the beautiful procedure of Balasubramanian and
Kraus \cite{balakraus}. Somewhat surprisingly, we find that it is
finite for all values of the parameter, which may be construed as a hint
that the duality is indeed valid even when the CTCs are present. However, it is
only a hint. If it turns out that the supergravity solutions with CTC's are simply not physical,
it would clearly be very interesting to identify the bound on the parameter
directly in ${\cal N}=4$ super-Yang-Mills theory.
Perhaps it represents a kind of unitary bound as, for example,
in the case of the BMPV black hole \cite{bmpv} where the onset of CTCs has been
argued to correspond to a unitary bound in the dual CFT \cite{herd}.

By construction, the D=5 solutions preserve two supersymmetries,
or 1/4 of the D=5 supersymmetry \cite{gaungut}. Now it is expected
that the full supersymmetric D=5 supergravity theory can be obtained via consistent
truncation of the Kaluza-Klein reduction of type IIB supergravity on a five-sphere
(this was only shown for the bosonic sector in \cite{cejm}), and hence
it is expected that the uplifted solutions will preserve at least two
supersymmetries or 1/16 of type IIB supergravity. We show here that they preserve
precisely two supersymmetries and present explicit expressions for the Killing spinors.
This is in marked contrast to the G\"odel solution of minimal ungauged supergravity
in D=5, which preserves eight supersymmetries in D=5 and yet, as remarked above,
twenty supersymmetries when uplifted to obtain a solution of D=11 supergravity.

Our method of determining the supersymmetry, mostly using a computer algebraic package, is
easily adapted to determine the preserved type IIB supersymmetry of any uplifted
solution of minimal gauged supergravity in D=5. Recently, black hole solutions
that are asymptotically globally $AdS_5$ were found \cite{gutreall} (similar solutions
were found for non-minimal D=5 gauged supergravity in \cite{gutreall2}),
and we show here that they also preserve just two supersymmetries when uplifted
to get type IIB solutions. This is very important in seeking a microscopic
interpretation of the entropy of these black holes: our result implies that
the entropy must correspond to states in ${\cal N}=4$ super-Yang-Mills theory preserving
precisely 1/16 supersymmetry.

The plan of the paper is as follows. Section 2 analyses the D=5 solution and
calculates the energy-momentum tensor. Section 3 analyses the uplifted
type IIB solution and determines the amount of preserved supersymmetry.
Analogous results for the uplifted $AdS_5$ black hole solutions are presented in
section 4. The paper contains three appendices.

\section{The five-dimensional solution}

We start with the family of solutions of minimal five-dimensional gauged supergravity
found in Ref.\cite{gaungut} whose metric, written in new co-ordinates, is
\bea
\label{themetric}
ds^2_5 = - \left( dt + \frac{r^2}{2l} \sigma_3^L
+ \frac{f r^2}{V(r)} \sigma_1^L \right)^2
+ \frac{dr^2}{V(r)} + \frac{r^2}{4} \left[ (\sigma_1^L)^2
+ (\sigma_2^L)^2 + V(r)\, (\sigma_3^L)^2 \right]
\eea
where $V(r) = 1 + r^2/l^2$ and the right-invariant
1-forms on the three-sphere, $\sigma_i^L$, are given by\footnote{These are dual to right-invariant
vector fields that generate left actions, and hence the superscript $L$.}
\begin{eqnarray}
\label{sigmadefs}
\sigma_1^L &=& \sin\phi d\theta - \sin\theta \cos\phi d\psi, \cr
\sigma_2^L &=& \cos\phi d\theta + \sin\theta \sin\phi d\psi, \cr
\sigma_3^L &=& d\phi + \cos\theta d\psi \ .
\end{eqnarray}
(Note that $\chi$, ${\cal F}_1, {\cal F}_2$
of \cite{gaungut} are related to $l, f$ via $l={\sqrt {12}}/\chi$ and
$f={\cal F}_1/4$ with ${\cal F}_2=0$.)
This metric is supported by a $U(1)$ gauge field whose field strength
is given by:
\begin{equation}
\label{gaugefield}
F^{(2)}=\frac{\sqrt 3}{2}f d\left(\frac{r^2}{V}\sigma^L_1\right)=
\frac{\sqrt 3}{2}f \left(\frac{2r}{V^2}dr\wedge\sigma^L_1-
\frac{r^2}{V}\sigma^L_2 \wedge \sigma^L_3\right) \ .
\end{equation}
Note that the Killing vector $\partial_t$ is timelike and that there are
additional Killing vectors that generate an $SU(2)_R$ group of isometries.
The solution preserves two supersymmetries, i.e. 1/4 of the
supersymmetry.

It is easy to see that when $f=0$, the metric (\ref{themetric})
is just that of $AdS_5$. To see this, set $f=0$ and
introduce the new co-ordinate
\be\label{coordchange}
\tilde{\phi}=\phi-2t/l
\ee
so that the metric simplifies to
\begin{equation}
\label{globalads}
ds^2 = - V(r)dt^2 + \frac{dr^2}{V(r)} +
\frac{r^2}{4} \left[ (\tilde{\sigma}_1^L)^2 + (\tilde{\sigma}_2^L)^2
  + (\tilde{\sigma}_3^L)^2
\right]
\end{equation}
where $\tilde{\sigma}_i^L$ are as in \p{sigmadefs} with $\phi$ replaced
with $\tilde{\phi}$.  This is $AdS_5$ in global coordinates. Thus the real
parameter $f$ is a measure of the deformation of the metric away from
that of $AdS_5$.  Note that performing the same coordinate change when
$f\ne 0$ leads to time-dependent metric components.

It was shown in \cite{gaungut} that the spacetime has
closed time-like curves when $f^2l^2$ is large.
It was shown in \cite{Behrndt:2003gc}, more precisely, that the metric has closed
time-like curves when $f^2>1/(4l^2)$, and that they are absent when $f^2\le 1/(4l^2)$.
Moreover, the closed time-like curves go through all points.

Next observe that the metric is locally asymptotically
$AdS_5$. This can be seen by evaluating the Riemann tensor
$R_{ab}^{~~cd}$. One finds:
\begin{equation}
\label{asymriemann}
R_{ab}^{~~cd} = - \frac{1}{l^2} (\delta_a^{~c} \delta_b^{~d} - \delta_a^{~d}
\delta_b^{~c}) + {\cal O}(\frac{1}{r})
\end{equation}
in the limit as $r \rightarrow \infty$.
Thus one expects on general grounds that the
holographic dual of the supergravity (and string) theory on this
background should be dual to four-dimensional ${\cal N}=4$ $SU(N)$
SYM on the boundary of this geometry.  The non-vanishing
$U(1)$ gauge-fields indicate that there are non-vanishing $R$-symmetry currents in the dual theory. This was also discussed in \cite{BK}.

We note that
\bea\label{dualfieldst}
*_{(5)} F^{(2)}=2\sqrt{3} f\Big[
&&\frac{r^2}{4V} dt\wedge\sigma_2^L\wedge\sigma_3^L
-\frac{r}{2V^2} dt\wedge dr\wedge\sigma_1^L\nn
&&-\frac{r^3}{4lV^2} dr\wedge \sigma_1^L\wedge\sigma_3^L
+\frac{fr^4}{4V^2} \sigma_1^L\wedge \sigma_2^L\wedge\sigma_3^L\Big] \ .
\eea
In particular if we integrate this over a surface $\Sigma$ defined
at fixed $r,t$ as $r\to\infty$ we get
\be
\int_\Sigma *_{(5)} F^{(2)}=8\sqrt{3}\pi^2f^2 l^4
\ee
indicating that the solution carries electric charge.

\subsection{The four-dimensional boundary geometry}
As $r\to \infty$ the metric
(\ref{themetric}) takes the form
\be\label{bdy metric}
ds^2=\frac{l^2}{r^2}dr^2+r^2(ds^2_{bdy})+{\cal O}(1)
\ee
where the boundary metric is defined as
\begin{equation}
\label{bdymetric}
ds^2_{bdy} = -\frac{1}{l}\, \left(dt + fl^2\sigma_1^L\right)\sigma_3^L
+ \frac{1}{4} \sigma_i^L \sigma_i^L \ .
\end{equation}
This is a regular metric and when $f=0$ we get the standard
metric on $\bR\times S^3$ after
employing the coordinate transformation \p{coordchange}.
On can calculate scalar curvature invariants and we find
that it has constant Ricci scalar curvature given by $R = 6$ and
$R_{\mu\nu}R^{\mu\nu}=R_{\mu\nu\rho\sigma}R^{\mu\nu\rho\sigma}=12$.
In particular we conclude that the Euler invariant,
$E\equiv R_{\mu\nu\rho\sigma}R^{\mu\nu\rho\sigma} - 4 R_{\mu\nu}R^{\mu\nu} +
R^2$, and the Weyl invariant
$C_{\mu\nu\rho\sigma}C^{\mu\nu\rho\sigma} \equiv
R_{\mu\nu\rho\sigma}R^{\mu\nu\rho\sigma} - 2 R_{\mu\nu}R^{\mu\nu} +
R^2/3$ both vanish.
Note that the boundary metric is not Einstein, and
since the Weyl tensor does not vanish, when $f\ne 0$,
it is not conformally flat.

Note that the killing vector $\partial/\partial
t$ is now null and that the metric maintains the right-$SU(2)_R$
isometries. Note also
that the condition for closed time-like curves on the boundary is the
same as that in the bulk.

\subsection{The energy-momentum tensor}

We now calculate the boundary energy-momentum tensor for this
system\footnote{After this paper was submitted to the arXive,
a revised version of  \cite{BK} appeared where this was
independently calculated \cite{BK}.}.
From \cite{balakraus} we have\footnote{Note that here we do not need any contributions
from the gauge-field of the type discussed in \cite{marika} to cancel any divergences.}
\begin{equation}
\label{emtone}
T_{\mu\nu} = \lim_{r\to\infty}
\frac{1}{8\pi G} r^2 \left[ \Theta_{\mu\nu} - \Theta \gamma_{\mu\nu}
-\frac{3}{l}\gamma_{\mu\nu} + \frac{l}{2} G_{\mu\nu} \right]
\end{equation}
where the extrinsic curvature
\begin{equation}
\label{extcurv}
\Theta_{\mu\nu} = \nabla_\mu \hat n_\nu
\end{equation}
is calculated with the unit normal $\hat n^\mu = (0, -V^{1/2}, 0, 0,
0)$ and the Einstein tensor $G_{\mu\nu} = R_{\mu\nu} - \frac{1}{2} R
\gamma_{\mu\nu}$ is calculated for the 4-d metric induced on a slice of
constant $r$. The factor of $r^2$ appearing in \p{emtone} is to
obtain the energy-momentum tensor on the boundary geometry \p{bdymetric}.
One finds
 \bea
8\pi G T&=&\frac{l}{2}dt^2-\frac{l^2}{8}dt(\sigma^L_3-24fl\sigma^L_1)
+\frac{l^3}{32}(1+80(fl)^2)(\sigma_1^L)^2\nn
&+&\frac{l^3}{32}(\sigma^L_2)^2
+\frac{l^3}{32}(1+64(fl)^2)(\sigma^L_3)^2-\frac{13fl^4}{8}\sigma^L_1\sigma^L_3 \ .
\eea
We have directly checked that this energy-momentum tensor
satisfies the conservation law $\nabla^\mu T_{\mu\nu} = 0$
with respect to the boundary metric \p{bdymetric}, as expected.
Furthermore, the trace with respect to the boundary metric
also vanishes, $T_\mu{}^\mu$=0, which is consistent with the vanishing of
the Euler and Weyl invariants of the boundary metric.
Finally, we observe that when $f=0$ we obtain the correct result for $AdS_5$.
In particular, setting $f=0$ and employing the
co-ordinate transformation \p{coordchange} we get,
\bea
8\pi GT&=&\frac{3l}{8}dt^2+\frac{l^3}{32}
\left[ ({\tilde\sigma}_1^L)^2 + (\tilde{\sigma}_2^L)^2
  + (\tilde{\sigma}_3^L)^2\right] \ .
\eea

To further analyse the EM tensor, we consider its components
with respect to an orthonormal frame for the boundary metric.
In particular, using the frame
\bea\label{oframe}
e^0&=&\frac{1}{l} dt+fl \sigma_1^L,\qquad
e^1=\frac{1}{2}\sigma_1^L,\nn
e^2&=&\frac{1}{2}\sigma_2^L,\qquad\qquad
e^3=\frac{1}{2}\sigma_3^L-\frac{1}{l}dt-fl\sigma_1^L
\eea
we find that the energy-momentum tensor has components
\be
8 \pi G T_{ab}=
     \frac{l^3}{8} \begin{bmatrix}
         3+x^2 & -x  & 0 & x^2  \\
         -x    & 1   & 0 & -3x  \\
         0     & 0   & 1 & 0   \\
         x^2   & -3x & 0 & 1+x^2
      \end{bmatrix}
\ee
where $x=8fl$.  The most striking feature of this energy-momentum
tensor is that it is not divergent for any value of the deformation
parameter.  Moreover, it does not undergo any particular
transformation as the CTC's appear when $x=4$.

To gain some further insight, we first recall (e.g. section 4.3 of
\cite{HE}) that any energy-momentum tensor in four dimensions is in
one of four canonical classes, depending on whether $T_{\mu}{}^{\nu}$
has one timelike eigenvector, a double null eigenvector, a triple null
eigenvector, or neither a time-like nor a null eigenvector,
respectively. Indeed for $x<x_c$, where $x_c\approx 2.515534$,
%
there is one time-like and three spacelike eigenvectors and we are in
the first class of \cite{HE}. Moreover, one can check that the weak
energy condition holds. For $x>x_c$ there are just two space-like
eigenvectors and we are in the fourth class of \cite{HE}. In \cite{HE}
it is stated that there are no observed fields with energy-momentum
tensors of this form, so this may be an indication that the
supergravity solutions are not physical, which would then also exclude
the solutions with CTC's. Of course the EM tensor above is the
expectation value in the quantum theory and moreover corresponds to
large 't Hooft coupling, so it seems quite plausible to us that the
solutions are physical all the way up to the onset of CTC's at $x=4$.

We should also point out that there are some well known ambiguities in the energy-momentum
tensor computed using the counterterm subtraction method of \cite{balakraus}.
Firstly, one is free to
add the following local gravitational counterterms to the boundary action:
\begin{equation}\label{gravct}
\frac{l^3}{8\pi G} \int_{\partial {\cal M}_r} d^4x \, \sqrt{-\gamma} \left[\alpha R^2
+ \beta R^{\mu\nu} R_{\mu\nu} + \gamma R^{\mu\nu\rho\sigma}
R_{\mu\nu\rho\sigma} \right],
\end{equation}
where $\alpha,\beta$ and $\gamma$ are constants.
These terms contribute \cite{Birrell:ix}
\begin{equation}
\label{extratermsone}
{\cal H}_{\mu\nu} = \frac{l^3}{4\pi G} \left[ \alpha \, {}^{(1)}\!H_{\mu\nu} + \beta
\, {}^{(2)}\!H_{\mu\nu} + \gamma \, H_{\mu\nu} \right]
\end{equation}
to the boundary EM tensor. We have presented the explicit expressions of
these quantities for our solution in appendix C. The fact that the Gauss-Bonnet
combination $R_{\mu\nu\rho\sigma} R^{\mu\nu\rho\sigma} +
R^2 -4 R_{\mu\nu} R^{\mu\nu}$ is topological in four-dimensions implies that
$H_{\mu\nu} = - {}^{(1)}\!H_{\mu\nu} +4 \, {}^{(2)}\!H_{\mu\nu}$.
Therefore we may always set one of $\alpha, \beta,
\gamma$ to zero, say $\gamma =0$. If the boundary metric
$\gamma_{\mu\nu}$ is conformally flat, $C_{\mu\nu\rho\sigma}=0$, we
further have ${}^{(2)}\!H_{\mu\nu} = (1/3) {}^{(1)}\!H_{\mu\nu}$ which
reduces the ambiguity down to a single parameter. Usually this
remaining ambiguity is fixed by demanding that the EM tensor of the
boundary CFT does not have any trace anomaly proportional to $\square
R$. If $\gamma_{\mu\nu}$ is not conformally flat ${\cal
H}_\mu{}^\mu=0$ requires $3\alpha +\beta =0$, which then leaves a single
free parameter.

Besides the finite gravitational counterterms,
there is also the following counterterm depending on the gauge field:
\begin{equation}
\label{emcounterterm}
-\frac{l\delta}{8\pi G} \int_{\partial {\cal M}_r} d^4x \, \sqrt{-\gamma} F^{(2)}_{\mu\nu}
F^{(2)\mu\nu}
\end{equation}
where $F^{(2)}$ is the two-form field-strength at fixed $r$ and $\delta$ is
another constant. These and related contributions have been considered recently
in a different context in \cite{LiuSabra,Caldarelli:2004ig}.
We have also listed the
contributions of these finite
counterterms to the EM tensor of our boundary CFT in appendix C.

\subsection{Conserved Quantities}

Having obtained the energy-momentum tensor we can determine the conserved
charges corresponding to the Killing vectors of the boundary metric as explained in
\cite{balakraus}. The general prescription is to first pick a spacelike 3-surface $\Sigma$
on the boundary with metric $\sigma_{ab}$. Let $u^\mu$ be a timelike
(not necessarily Killing) vector normal to $\Sigma$ and let $\xi^\mu$
be a Killing vector of the boundary metric. Then the conserved
quantity associated to this Killing vector is:
\begin{equation}
\label{consquant}
Q_{\xi} = \int_\Sigma d^3x \, \sqrt{\sigma} ( u^\mu \xi^\nu
T_{\mu\nu}) \ .
\end{equation}

Consider the surface $\Sigma$ defined by $t$=constant. We
restrict to the case of no CTC's, $4f^2l^2<1$, in which case $\Sigma$
is indeed spacelike. The unit normal $u$, written as a one-form,
is equal to $-dt/(l \sqrt{1-4 l^2 f^2})$.
A straightforward calculation reveals that the conserved quantities associated with the
$SU(2)$ Killing-vectors $\xi^R_i$ all vanish (for explicit expressions for $\xi^R_i$ see, e.g.
appendix A of \cite{gauntharv}).
Similarly, the conserved quantity for the null Killing vector $\xi=\frac{\partial}{\partial t}$
is given by
\begin{equation}
Q_{\frac{\partial}{\partial t}} =\frac{1}{8\pi G}\frac{l^2 (3 +32 l^2 f^2) \pi^2}{4} \ .
\end{equation}
It is interesting to note that when written
in the co-ordinates \p{coordchange} the null Killing-vector becomes
$\frac{\partial}{\partial t}-(2/l)\xi^L_3$,
where $\xi^L_3=\frac{\partial}{\partial \phi}$. Now when $f=0$, the five-dimensional
solution is $AdS_5$, and both $\frac{\partial}{\partial t}$ (in the new co-ordinates)
and $\xi^L_3$ are Killing-vectors, with the former generating global time
translations. Thus when $f=0$ the conserved quantity $Q_{\frac{\partial}{\partial t}}$ is
simply the linear combination of charges $M-(1/l)(J_1+J_2)$,
where $M$ is the mass and $J_i$ are the two
$SO(4)$ Casimirs. Since $J_i=0$ for global $AdS_5$, when $f=0$ we recover
the mass of global $AdS_5$, which is
naturally interpreted as the Casimir energy of ${\cal N}=4$ super-symmetric Yang-Mills theory
on $\bR\times S^3$ \cite{balakraus}. When $f\ne 0$, neither $M$ nor $(J_1+J_2)$ are
separately conserved quantities, but the linear combination $Q_{\frac{\partial}{\partial t}}$ is.

Finally, we comment on the effect of the counterterms \p{gravct} and \p{emcounterterm}
on the conserved charges.
We find that the gauge-field counterterm \p{emcounterterm} does not affect the $Q_{\xi^R_i}$ or
$Q_{\frac{\partial}{\partial t}}$. The gravitational counter-terms \p{gravct} leave the charges
$Q_{\xi^R_i}$ invariant, but they do modify $Q_{\frac{\partial}{\partial t}}$. In particular, we
find, when $\gamma=0$,
\begin{equation}
\Delta Q_{\frac{\partial}{\partial t}} =-\frac{1}{8\pi G}24 l^2(3\alpha+\beta) \pi^2 \ .
\end{equation}
It is interesting to note that this vanishes when $3\alpha +\beta =0$, which is precisely
the condition required for the EM tensor not to have any trace anomaly proportional to $\square R$.

\section{The IIB solution and its supersymmetry}

Any bosonic solution of the minimal gauged supergravity in D=5 can be uplifted to
obtain a solution of ten-dimensional type IIB supergravity, with
the only non-vanishing fields being the metric and the self-dual five-form
using\footnote{Note a typo in the factor appearing in the Chern-Simons term
in \cite{cejm}: see also \cite{cveticetal}.}
the formulae in \cite{cejm}. Such uplifted solutions are expected to preserve
at least as many supersymmetries as the solution in D=5.

Specifically,  starting from a D=5 solution with metric $ds_5^2$ and field strength
$F^{(2)}= dA$, the uplifted type IIB supergravity solution is given by:
\begin{eqnarray}
\label{tendlift}
ds^2_{10} &=&
ds^2_5+ l^2 \sum_{i=1}^3 \left[ (d\mu_i)^2 + \mu_i^2
\left( d\xi_i + \frac{2}{l\sqrt{3}} A \right)^2 \right], \cr
F^{(5)} &=& (1+ *_{(10)}) \left[
-\frac{4}{l} {\rm{vol}_{(5)}} + \frac{{\it{l}}^2}{\sqrt{3}}
\sum_{i=1}^3 d(\mu_i^2) \wedge d\xi_i \wedge *_{(5)} F^{(2)} \right]
\end{eqnarray}
where $\mu_1 = \sin\alpha$, $\mu_2 = \cos\alpha\, \sin\beta$,
$\mu_3 = \cos\alpha\, \cos\beta$ with $0\le \alpha\le \pi/2$,
$0\le\beta\le \pi/2$, $0\le \xi_i\le 2\pi$
and together they parametrise $S^5$. Note that we define the Hodge star of a
$p$-form $\omega$ in $n$-dimensions as
$*_{(n)}\omega_{i_1\dots i_{n-p}}=\frac{1}{p!}\e_{i_1\dots i_{n-p}}{}^{j_1\dots
j_p}\omega_{j_1\dots j_p}$, with $\e_{0123456789}=1$ and $\e_{01234}=1$ in an
orthonormal frame. Finally, $\rm{vol}_{(5)}=\e$ is the volume five-form of
the metric \p{themetric}.

Any uplifted D=5 solution satisfies the type IIB Einstein equations
\be
R_{mn} - \frac{1}{96}F^{(5)}_{mp_1p_2p_3p_4} {
F^{(5)}}_n{}^{p_1p_2p_3p_4} = 0 \ .
\ee
The five-form is obviously self-dual, $F^{(5)}=*_{(10)}F^{(5)}$,
and one can show that it also satisfies the Bianchi identity
\be
dF^{(5)}=0 \ .
\ee
We have directly checked that uplifting the solution \p{themetric}, \p{gaugefield}
solves these IIB equations of motion.

In our conventions the conditions for preserved IIB supersymmetry are given
by
\bea
{\hat \nabla}_m\e &\equiv&
D_m \epsilon + \frac{i}{1920}
\Gamma^{n_1n_2n_3n_4n_5}\Gamma_{m}F^{(5)}_{n_1n_2n_3n_4n_5} \epsilon
\nn
&=&D_m \epsilon + \frac{i}{192}
\Gamma^{n_1n_2n_3n_4}F^{(5)}_{mn_1n_2n_3n_4} \epsilon=0 \ .
\eea
Here $\e=\e^1+i\e^2$ and $\e^1$, $\e^2$ are both Majorana-Weyl spinors with
\be
\Gamma_{11} \epsilon = -\epsilon
\ee
where $\Gamma_{11} = \Gamma_0 \Gamma_1 \dots \Gamma_9$.
Note that the integrability conditions for supersymmetry,
$[{\hat \nabla}_m, {\hat \nabla}_n] \, \epsilon = 0$, imply that
\bea\label{intconds}
&&\big[ R_{mn s_1 s_2}-\frac{1}{48} F^{(5)}{}_{m s_1 r_1 r_2 r_3}
F^{(5)}{}_{n s_2}{}^{r_1 r_2 r_3} \big] \Gamma^{s_1 s_2} \epsilon
\nn
&&+\big[ \frac{i}{24}\nabla_{[m} F^{(5)}{}_{n] s_1 s_2 s_3 s_4}
+\frac{1}{96} F^{(5)}{}_{m n r_1 r_2 s_1} F^{(5)}{}^{r_1 r_2}{}_{s_2 s_3 s_4} \big]
\Gamma^{s_1 s_2 s_3 s_4} \epsilon =0
\eea
where in obtaining this expression we have made use of the identity
\be
F^{(5)}{}^{r_1 r_2 r_3}{}_{m_1 [m_2}F^{(5)}{}_{m_3 m_4] r_1 r_2 r_3} =0 \ .
\ee

To calculate the preserved supersymmetries we will use the following,
slightly non-obvious, frame. We take $e^0,e^1,e^2,e^3,e^4$ to be a frame
for the five-dimensional solution given by
\begin{eqnarray}
\label{vbsone}
e^0 &=& dt + \frac{r^2}{2l} \sigma_3^L + \frac{fr^2}{V} \sigma_1^L,\qquad\qquad\qquad
e^1 = \frac{1}{V^{1/2}} \, dr,
\nn
e^2 &=& \frac{r}{2}\sigma_1^L,\qquad\qquad
e^3 = \frac{r}{2} \sigma_2^L,\qquad\qquad
e^4 = \frac{r}{2} V^{1/2} \, \sigma_3^L,
\end{eqnarray}
and supplement this with
\begin{eqnarray}
\label{vbs}
e^5 &=& l \, d\alpha,\qquad\qquad
e^6 = l \, \cos\alpha \, d\beta,
\cr
e^7 &=&  l \, \sin\alpha \, \cos\alpha \, [ d\xi_1 - \sin^2\beta \,
d\xi_2 - \cos^2\beta \, d\xi_3 ],
\cr
e^8 &=& l \, \cos\alpha \, \sin\beta \, \cos\beta \, [ \, d\xi_2 -
  d\xi_3 ],
\cr
e^9
&=& -\frac{2}{\sqrt 3}A\, -l\, \sin^2\alpha \, d\xi_1
- l \, \cos^2\alpha \, ( \sin^2\beta \, d\xi_2 + \cos^2\beta \, d\xi_3)
\end{eqnarray}
where $A=({\sqrt 3}/2)(fr^2/V)\sigma^L_1$ for the solution under consideration.
Thus we have
\begin{eqnarray}\label{effexp}
F^{(5)}
&=&-4 l^{-1} [ e^0
\wedge e^1\wedge e^2 \wedge e^3 \wedge e^4 + e^5 \wedge e^6
\wedge {e}^7 \wedge {e}^8 \wedge {e}^9 ]
\cr
&& +\frac{2}{\sqrt 3}(e^5\wedge e^7+e^6\wedge e^8)\wedge(*_{(5)}F^{(2)}
-e^9\wedge F^{(2)})
\cr
&=&-4 l^{-1} [ e^0
\wedge e^1\wedge e^2 \wedge e^3 \wedge e^4 + e^5 \wedge e^6
\wedge {e}^7 \wedge {e}^8 \wedge {e}^9 ]
\cr
&& - 4 f V^{-3/2} \, (e^0 + {e}^9) \wedge (e^1 \wedge e^2 - e^3
\wedge e^4) \wedge (e^5 \wedge {e}^7  + e^6 \wedge {
e}^8) 
\end{eqnarray}
where the first expression is valid for a general solution of
minimal gauged supergravity.
The spin connection components for the solution \p{themetric}, \p{gaugefield}
are contained in the appendix.

Using a computer algebra package, the constraints on the Killing spinor
imposed by the integrability conditions can be obtained. In particular,
in the above orthonormal frame, we find the constraints
\bea
\label{eqn:alg}
\Gamma^{0149}\e&=&-i\e,\qquad \Gamma^{0239}\e=i\e\nn
\Gamma^{0579}\e&=&-i\e,\qquad \Gamma^{09}\e=\e \ .
\eea
Hence we observe that the solution preserves at most 1/16 of the supersymmetry,
or two supersymmetries. Now, since the solution was obtained by
uplifting a solution of five-dimensional minimal gauged supergravity
that preserved 2 supersymmetries, the ten-dimensional solution is expected to
preserve at least 2 supersymmetries and hence \p{eqn:alg} indicates that it
preserves precisely 2 supersymmetries. Indeed just using these constraints we can solve the
Killing spinor equation to obtain the Killing spinors in explicit
form.
We find the simple expression
\be
\epsilon = e^{\frac{i}{2}(\frac{3t}{\ell}-\xi_1 - \xi_2 - \xi_3)} \eta
\ee
where $\eta$ is a constant spinor, satisfying the projections
\p{eqn:alg}. Some useful formulae for verifying this result are presented in
an appendix.  Note that we have checked that the solution
obtained by changing the sign of the five-form also preserves two
supersymmetries. It is interesting to note that the projections are the same
projections for three orthogonally intersecting D3-branes with momentum along the common
intersection direction (the nine direction).

\section{Supersymmetry of uplifted $AdS_5$ black holes}

Asymptotically $AdS_5$ black hole solutions of minimal $D=5$ gauged supergravity with regular
horizons have recently been constructed in \cite{gutreall}. Here we uplift these
solutions to obtain solutions of type IIB supergravity and show that they preserve two
supersymmetries.

The metric of the five-dimensional solution is specified by
the funfbein given by
\bea
\label{eqn:fivebhviel}
e^0 &=& {\cal{F}} \big( dt + \Psi \sigma_3^L \big)
\nn
e^1 &=& {\cal{F}}^{-1} (1+{2 \omega^2 \over l^2}+{r^2 \over l^2})^{-{1 \over 2}} dr
\nn
e^2 &=& {r \over 2} \sigma_1^L
\nn
e^3 &=& {r \over 2} \sigma_2^L
\nn
e^4 &=& {r \over 2 l} \sqrt{l^2+2 \omega^2+r^2} \sigma_3^L
\eea
where
\be
{\cal{F}} = 1-{\omega^2 \over r^2}
\ee
and
\be
\Psi = -{\eta r^2 \over 2 l}\big( 1+ {2 \omega^2 \over r^2} +{3 \omega^4 \over 2r^2(r^2-\omega^2)}
\big)
\ee
where $\eta= \pm 1$ and $\omega$ is constant. The 2-form gauge potential is given by
\be
A = {\sqrt{3} \over 2} \big[ {\cal{F}} dt +{\eta \omega^4 \over 4 l r^2} \sigma_3^L \big] \ .
\ee

Hence the metric of the uplifted solution is obtained by taking the ten-dimensional
frame $e^0$, $e^1$, $e^2$, $e^3$ and $e^4$ given above in (\ref{eqn:fivebhviel})
together with $e^5,e^6,e^7,e^8,e^9$ given in \p{vbs}.
Using the first line of \p{effexp} we deduce that the self-dual
five-form field strength is given by
\bea
\label{eqn:bhfiveform}
F^{(5)} &=& -4 l^{-1} \big(e^0 \wedge e^1 \wedge e^2 \wedge e^3 \wedge e^4
+ e^5 \wedge e^6 \wedge e^7 \wedge e^8 \wedge e^9 \big)
\nn
&+& \big[ -{\eta \omega^4 \over l r^4} (e^0 \wedge e^1 \wedge e^4 -
e^2 \wedge e^3 \wedge e^9)+{\eta \omega^2 \over l r^4} (2r^2+\omega^2)
(e^0 \wedge e^2 \wedge e^3 - e^1 \wedge e^4 \wedge e^9)
\nn
&+& {2 \omega^2 \over l r^3} \sqrt{l^2+2 \omega^2+r^2}
(e^0 \wedge e^1 \wedge e^9 +e^2 \wedge e^3 \wedge e^4) \big]
\wedge (e^5 \wedge e^7 + e^6 \wedge e^8) \ .
\eea

As in the last section, by first examining the integrability conditions for supersymmetry
\p{intconds},
we find that they type IIB solution preserves exactly two supersymmetries. The explicit
form of the Killing spinors is
\be
\label{eqn:bhkillsp}
\epsilon = {\cal{F}}^{1 \over 2} e^{-{i \over 2} (\xi_1 + \xi_2 + \xi_3)} \epsilon_0
\ee
where $\epsilon_0$ is a constant spinor satisfying the constraints
\bea
\label{eqn:bhconstraint}
\Gamma^{0149}\e_0&=&i \eta \e_0,\qquad \Gamma^{0239}\e_0=-i \eta \e_0\nn
\Gamma^{0579}\e_0&=&-i\e,\qquad \Gamma^{09}\e_0=\e_0 \ .
\eea
Note that when $\eta=-1$ these are precisely
the same projections as \p{eqn:alg}.

\section*{Acknowledgements}
It is a pleasure to thank Per Kraus and Rob Myers for helpful discussions.

\appendix
\section{Formulae for Killing spinor calculation I}
For the deformation of $AdS_5\times S^5$ solution, the spin connection for the frame
\p{vbsone}, \p{vbs} can be calculated and used to obtain $D_\mu\e$. We find:
\begin{eqnarray}
&& \partial_t \epsilon + \frac{1}{2}\Big[ \frac{1}{l}(\Gamma_{14} -
\Gamma_{23}) +
\frac{2f}{V^{3/2}} (\Gamma_{12} - \Gamma_{34}) \Big] \epsilon,
\cr &&
\partial_{r} \epsilon - \frac{1}{2} \Big[ \frac{1}{lV^{1/2}} \Gamma_{04}
+\frac{2f}{V^2} (\Gamma_{29} +\Gamma_{02})
\Big] \epsilon,
\cr &&
\partial_\theta \epsilon + \frac{1}{2} \Big[
- \frac{r}{2l} ( \cos\phi\, \Gamma_{02} - \sin\phi \, \Gamma_{03})
- \frac{V^{1/2}}{2} [\cos\phi \, (\Gamma_{13} - \Gamma_{24}) + \sin\phi \,
(\Gamma_{12} + \Gamma_{34})]
\cr && ~~~~
+\frac{fr}{V^{3/2}} [ \cos\phi \,(\Gamma_{04} + \Gamma_{49}) + \sin\phi \, (\Gamma_{01} +\Gamma_{19})]
+ \frac{fr^2}{lV} \sin\phi \, (\Gamma_{14} -\Gamma_{23} - \Gamma_{57} - \Gamma_{68})
\Big] \epsilon,
\cr &&
\partial_\phi \epsilon + \frac{1}{2} \Big[- \frac{1}{2} \Gamma_{23} + \frac{rV^{1/2}}{2l} \Gamma_{01} -
\frac{V}{2} \Gamma_{14}
+\frac{fr^2}{lV^{3/2}}  (\Gamma_{12} -
\Gamma_{34}) - \frac{fr}{V} ( \Gamma_{03} + \Gamma_{39})
\Big] \epsilon,
\cr &&
\partial_\psi \epsilon + \frac{1}{2} \Big[ - \frac{1}{2} \cos\theta \, \Gamma_{23}
- \frac{r}{2l} \sin\theta \, ( \cos\phi \, \Gamma_{03} + \sin\phi \,
\Gamma_{02}) + \frac{rV^{1/2}}{2l} \cos\theta \, \Gamma_{01}
\cr && ~~~~
+ \frac{V^{1/2}}{2} [\sin\theta \, \cos\phi ( \Gamma_{34} +
\Gamma_{12})   - \sin\theta \, \sin\phi \,
(\Gamma_{13} - \Gamma_{24})] - \frac{V}{2} \cos\theta \, \Gamma_{14}
\cr && ~~~~
+\frac{fr}{V^{3/2}} \sin\theta \,
[ \cos\phi \, (-\Gamma_{19} - \Gamma_{01}) + \sin\phi \, (\Gamma_{04}
 + \Gamma_{49})]
\cr && ~~~~
- \frac{fr^2}{lV^{3/2}} \cos\theta \, (\Gamma_{34}
- \Gamma_{12}) + \frac{fr}{V} \cos\theta \, (-\Gamma_{39} -
\Gamma_{03})
\cr && ~~~~
+ \frac{fr^2}{lV} \sin\theta \, \cos\phi \, (\Gamma_{57} + \Gamma_{68}
- \Gamma_{14} + \Gamma_{23})
\Big] \epsilon,
\cr &&
\partial_\alpha \epsilon - \frac{1}{2} \Gamma_{79} \epsilon,
\cr &&
\partial_\beta \epsilon + \frac{1}{2} \Big[ \sin\alpha \, (\Gamma_{56}
 + \Gamma_{78}) - \cos\alpha \, \Gamma_{89} \Big] \epsilon,
\cr &&
\partial_{\xi_1} \epsilon - \frac{1}{2} \Big[  \cos\alpha \, (\cos\alpha \,
\Gamma_{57} - \sin\alpha \, \Gamma_{59})
+\frac{2fl}{V^{3/2}}
  \sin^2\alpha \,
( \Gamma_{12} - \Gamma_{34})
\Big] \epsilon,
\cr &&
\partial_{\xi_2} \epsilon +\frac{1}{2} \Big[
- \sin\alpha \, \sin\beta \,
\left( \sin\alpha \, \sin \beta \, \Gamma_{57} - \cos\beta \,
  \Gamma_{58} + \cos\alpha \, \sin\beta \, \Gamma_{59} \right)
\cr && ~~~~
+ \cos\beta
\, \left( \sin\alpha \, \sin\beta \, \Gamma_{67} - \cos\beta \,
  \Gamma_{68} + \cos\alpha \, \sin\beta \, \Gamma_{69} \right)
\cr && ~~~~
+\frac{2fl}{V^{3/2}}
  \cos^2\alpha \,
\sin^2\beta ( \Gamma_{34} - \Gamma_{12} )
\Big]\epsilon,
\cr &&
\partial_{\xi_3} \epsilon +\frac{1}{2} \Big[
- \sin\alpha \, \cos\beta
\, \left( \sin\alpha \, \cos\beta \, \Gamma_{57} + \sin\beta \,
\Gamma_{58} + \cos\alpha \, \cos\beta \, \Gamma_{59} \right)
\cr && ~~~~
- \sin\beta \, \left( \sin\alpha \, \cos\beta \, \Gamma_{67} +
\sin\beta \, \Gamma_{68} + \cos\alpha \, \cos\beta \, \Gamma_{69}
\right)
\cr && ~~~~
+\frac{2fl}{V^{3/2}}  \cos^2\alpha \,\cos^2\beta \, (\Gamma_{34} - \Gamma_{12})
\Big] \epsilon \ .
\end{eqnarray}
In determining the Killing spinors it is also useful to note
%
%
\begin{eqnarray}
\frac{i}{1920}\Gamma^{n_1n_2n_3n_4n_5}F^{(5)}_{n_1n_2n_3n_4n_5} &=&
\frac{i}{4}\Big[ \frac{1}{l}\Gamma_{01234}
+ \frac{f}{V^{3/2}}\Gamma_0 (\Gamma_{57}+\Gamma_{68})(\Gamma_{12}-\Gamma_{34})
\Big](1+\Gamma_{11}) \ . \nn
\end{eqnarray}

\section{Formulae for Killing spinor calculation II}

The formula for $D_\mu \epsilon$ for the frame
({\ref{eqn:fivebhviel}}), \p{vbs} of the uplifted AdS
black holes is given by

\bea
\label{eqn:bhspincon}
\partial_t \epsilon &+& {1 \over 2} \big[ {\sqrt{l^2+r^2+2 \omega^2}} {\omega^2 \over l r^3} {\cal{F}} (\Gamma_{01}
+\Gamma_{19})
-{\eta \over l} {\cal{F}}^2 \Gamma_{14}
+{(r^2+\omega^2) \eta \over r^2 l} {\cal{F}} \Gamma_{23}
\nn
&-& {{\cal{F}} \over l} (\Gamma_{57} + \Gamma_{68}) \big] \epsilon \ ,
\nn
\partial_r \epsilon &+&{1 \over 2} \big[- {\eta (4 \omega^2 r^2 + \omega^4 -2r^4)
\over 2 r^4 \sqrt{l^2+r^2+2 \omega^2} {\cal{F}}} \Gamma_{04}
+{\omega^2 \over r^3 {\cal{F}}} \Gamma_{09} -{\eta \omega^2 (2r^2+\omega^2) \over 2 r^4 \sqrt{l^2+r^2+2 \omega^2}
{\cal{F}}} \Gamma_{49} \big] \epsilon \ ,
\nn
\partial_\theta \epsilon &+& {1 \over 2} \big[{\eta \over 4 r^3 l}  (-2r^4 -2 r^2 \omega^2 + \omega^4)
(- \cos \phi \Gamma_{02} + \sin \phi \Gamma_{03})
\nn
&-&{1 \over 2 l}  {\cal{F}} \sqrt{l^2+r^2+2 \omega^2} (\sin \phi \Gamma_{12}
+ \cos \phi \Gamma_{13})
\nn
&+&{1 \over 2l}  \sqrt{l^2+r^2+2 \omega^2} (\cos \phi \Gamma_{24} - \sin \phi \Gamma_{34})
-{\eta \omega^4  \over 4 r^3 l} (\cos \phi \Gamma_{29} - \sin \phi \Gamma_{39})
\big] \epsilon \ ,
\nn
\partial_\phi \epsilon &+&{1 \over 2} \big[ {\eta \over 4 r^5 l^2} \sqrt{l^2+r^2+2 \omega^2} (-3 \omega^4 r^2
+\omega^6-2r^6) \Gamma_{01}
\nn
&-&{1 \over 2 r^2 l^2}(r^2 \omega^2 + \omega^4 + r^4 + r^2 l^2- \omega^2 l^2) \Gamma_{14}
-{\eta \omega^4 \over 4 r^3 l^2} \sqrt{l^2+r^2+2 \omega^2} {\cal{F}} \Gamma_{19}
\nn
&+&{1 \over 2 r^4 l^2}(\omega^6-r^4 l^2) \Gamma_{23}-{\eta \omega^4 \over 4 l^2 r^2}
(\Gamma_{57}+\Gamma_{68}) \big] \epsilon
\nn
\partial_\psi \epsilon &+& {1 \over 2} \big[ {\eta \over 4 r^5 l^2} \cos \theta \sqrt{l^2+r^2+2 \omega^2}
(-3 \omega^4 r^2+ \omega^6 -2 r^6) \Gamma_{01}
\nn
&-& {\eta \over 4 r^3 l} \sin \theta  (-2 r^4 -2 r^2 \omega^2 +\omega^4) (\sin \phi \Gamma_{02}
+ \cos \phi \Gamma_{03})
\nn
&+&{1 \over 2l} \sin \theta \sqrt{l^2+r^2+2 \omega^2} {\cal{F}}
(\cos \phi \Gamma_{12}- \sin \phi \Gamma_{13})
\nn
&-& {1 \over 2 r^2 l^2} \cos \theta (r^2 \omega^2 + \omega^4 + r^4 +r^2 l^2 - \omega^2 l^2) \Gamma_{14}
-{\eta \over 4 l^2 r^2} \omega^4 \cos \theta  (\Gamma_{57}+\Gamma_{68})
\nn
&-&{\eta \over 4 r^3 l^2} \cos \theta \sqrt{l^2+r^2+2 \omega^2} \omega^4 {\cal{F}} \Gamma_{19}
+{1 \over 2 r^4 l^2} \cos \theta (\omega^6-r^4 l^2) \Gamma_{23}
\nn
&+& {1 \over 2l} \sin \theta  \sqrt{l^2+r^2+2 \omega^2} ( \sin \phi \Gamma_{24}+ \cos \phi \Gamma_{34})
-{\eta \over 4 r^3 l} \sin \theta \omega^4 (\sin \phi \Gamma_{29} + \cos \phi \Gamma_{39})
\big] \epsilon
\nn
\partial_\alpha \epsilon &-& {1 \over 2} \Gamma_{79} \epsilon
\nn
\partial_\beta \epsilon &+&{1 \over 2} \big[\sin \alpha (\Gamma_{56}+\Gamma_{78}) - \cos \alpha
\Gamma_{89} \big] \epsilon
\nn
\partial_{\xi_1} \epsilon &+& {1 \over 2} \big[ -{\omega^2 \over r^3} \sin^2 \alpha \sqrt{l^2+r^2
+2 \omega^2} \Gamma_{01}-{\eta \over 2 r^4} \omega^2 \sin^2 \alpha (2r^2+\omega^2) \Gamma_{14}
\nn
&+&{\eta \over 2 r^4} \omega^4 \sin^2 \alpha  \Gamma_{23} - \cos^2 \alpha \Gamma_{57}+ \cos
\alpha \sin \alpha \Gamma_{59} \big] \epsilon
\nn
\partial_{\xi_2} \epsilon &+&{1 \over 2}  \big[ -{\omega^2 \over r^3} \cos^2 \alpha \sin^2 \beta \sqrt{
l^2+r^2+ 2 \omega^2} \Gamma_{01} -{\eta \omega ^2 \over 2 r^4} \cos^2 \alpha \sin ^2 \beta (2r^2+\omega^2)
\Gamma_{14}
\nn
&+& {\eta \omega^4 \over 2 r^4} \cos^2 \alpha \sin^2 \beta \Gamma_{23}
-\sin^2 \beta \sin^2 \alpha \Gamma_{57} + \sin \beta \cos \beta \sin \alpha (\Gamma_{58}+ \Gamma_{67})
\nn
&-& \cos \alpha \sin \alpha \sin^2 \beta \Gamma_{59} - \cos^2 \beta \Gamma_{68}
+ \cos \alpha \sin \beta \cos \beta \Gamma_{69} \big] \epsilon
\nn
\partial_{\xi_3} \epsilon &+&{1 \over 2} \big[ -{l \omega^2 \over r^3} \cos^2 \alpha \cos^2 \beta
\sqrt{l^2+r^2+2 \omega^2} \Gamma_{01} -{\eta l \omega^2 \over 2 r^4} (2r^2+\omega^2) \cos^2 \alpha
\cos^2 \beta \Gamma_{14}
\nn
&+&{\eta l \omega^4 \over 2 r^4} \cos^2 \alpha \cos^2 \beta \Gamma_{23}
-l \cos^2 \beta \sin^2 \alpha \Gamma_{57}-l \sin \alpha \sin \beta \cos \beta (\Gamma_{58}+\Gamma_{67})
\nn
&-& l \cos \alpha \sin \alpha \cos^2 \beta \Gamma_{59} - l \sin^2 \beta \Gamma_{68}
-l \cos \alpha \sin \beta \cos \beta \Gamma_{69} \big] \epsilon \ .
\nn
\eea

\section{Ambiguities in the energy-momentum tensor}
Using signature $(-,+,+,+)$ and standard conventions for
curvature tensors, the expressions for the tensors obtained from
the variation of \p{gravct} and given in \cite{Birrell:ix}, become
\begin{eqnarray}
{}^{(1)}\!H_{\mu\nu} &=& 2 \nabla_\mu \nabla_\nu R - 2g_{\mu\nu} \,
\square R + \frac{1}{2} g_{\mu\nu} R^2 - 2 R \, R_{\mu\nu}, \cr
{}^{(2)}\!H_{\mu\nu} &=&  \nabla_\mu \nabla_\nu R - \frac{1}{2}
g_{\mu\nu} \, \square R - \square R_{\mu\nu} + \frac{1}{2}g_{\mu\nu}
R^{\alpha\beta} R_{\alpha\beta} - 2 R^{\alpha\beta} R_{\alpha \mu
  \beta \nu}, \cr
H_{\mu\nu} &=& \frac{1}{2}g_{\mu\nu} R^{\alpha\beta\gamma\delta}
R_{\alpha\beta\gamma\delta} - 2 R_{\mu\alpha\beta\gamma}
R_\nu^{~\alpha\beta\gamma} - 4\, \square R_{\mu\nu} + 2\nabla_\mu
\nabla_\nu R \cr && ~ + 4 R_{\mu\alpha} R^\alpha_{~\nu} - 4
R^{\alpha\beta} R_{\alpha \mu \beta \nu} \ .
\end{eqnarray}
These are each conserved and satisfy  $H = -{}^{(1)}H + 4\,{}^{(2)}H$.
For the deformed $AdS_5$ background one finds that
${}^{(1)}H_{\mu\nu}$ and ${}^{(2)}H_{\mu\nu}$ can be written in terms of the
orthonormal frame \p{oframe} as
\begin{equation}
{}^{(1)}H_{ab} =  -6 \begin{bmatrix}
         3 & -x  & 0 & 0  \\
         -x    & 1   & 0 & -x  \\
         0     & 0   & 1 & 0   \\
         0   & -x & 0 & 1
      \end{bmatrix}
\end{equation}
\begin{equation}
{}^{(2)}H_{ab} =  -2 \begin{bmatrix}
         3+\frac{3}{4}x^2 & -x   &  0  & \frac{3}{4}x^2  \\
         -x              & 1    &  0  & -x  \\
         0               & 0    &  1  & 0   \\
         \frac{3}{4}x^2   & -x  &  0  & 1+\frac{3}{4}x^2
      \end{bmatrix}
\end{equation}
%

The contribution coming from \p{emcounterterm} is also easily calculated.
The contribution coming from $\gamma_{ab} F^2$ vanishes while the contribution
coming from $F_{ac}F_b{}^c$ gives
\begin{equation}
 -\frac{1}{8\pi G}\frac{3l^3\delta}{4} \begin{bmatrix}
         x^2 & 0  & 0 & x^2  \\
         0   & 0  & 0 & 0  \\
         0   & 0  & 0 & 0   \\
         x^2 & 0  & 0 & x^2
      \end{bmatrix}
\end{equation}

\end{document}